\documentclass[amsmath,amssymb,12pt,superscriptaddress,nofootinbib]{revtex4}
\usepackage[dvipdfmx]{graphicx}
\usepackage{bm}
\usepackage{color}
\usepackage{soul} 
\usepackage{caption} 

\sethlcolor{yellow}

\newcommand{\dd}{\mathrm{d}}

\def\bm#1{\mbox{\boldmath ${#1}$}} 

\definecolor{Red}{rgb}{1,0,0} 

\definecolor{Blue}{rgb}{0,0,1} 

\definecolor{Green}{rgb}{0,1,0}

\begin{document}
\baselineskip5.5mm
\thispagestyle{empty}

{\baselineskip0pt
\leftline{\baselineskip14pt\sl\vbox to0pt{
               \hbox{\it Division of Particle and Astrophysical Science}
              \hbox{\it Nagoya University}
                             \vss}}
\rightline{\baselineskip16pt\rm\vbox to20pt{
\vss}}
}

\author{Kengo Iwata}\email{iwata@gravity.phys.nagoya-u.ac.jp}
\affiliation{ 
Gravity and Particle Cosmology Group,
Division of Particle and Astrophysical Science,
Graduate School of Science, Nagoya University, 
Nagoya 464-8602, Japan
}

\author{Chul-Moon~Yoo}\email{yoo@gravity.phys.nagoya-u.ac.jp}
\affiliation{
Gravity and Particle Cosmology Group,
Division of Particle and Astrophysical Science,
Graduate School of Science, Nagoya University, 
Nagoya 464-8602, Japan
}

\vskip2cm
\title{Another approach to test gravity around a black hole}

\begin{abstract}
\vskip0.5cm
Pulsars orbiting around the black hole at our galactic center 
provide us a unique testing site for gravity. 
In this work, we propose an approach to probe the gravity around the black hole 
introducing two phenomenological parameters which 
characterize deviation from the vacuum Einstein theory. 
The two phenomenological parameters are associated with 
the energy momentum tensor in the framework of the Einstein theory. 
Therefore, our approach can be regarded as the complement 
to the parametrized post-Newtonian framework 
in which phenomenological parameters are introduced for 
deviation of gravitational theories from general relativity. 
In our formulation, we take the possibility of existence of 
a relativistic and exotic matter component into account. 
Since the pulsars can be regarded as test particles, as the first step, 
we consider geodesic motion in the system 
composed of a central black hole and a perfect fluid 
whose distribution is static and spherically symmetric. 
It is found that 
the mass density of the fluid and a parameter of the equation of state 
can be determined with precision with $0.1\%$ 
if the density on the pulsar orbit is larger than $10^{-9}~{\rm g/cm^3}$.

\end{abstract}

\maketitle
\pagebreak

\section{Introduction}
\label{sec:Intro}

After the invention of general relativity, 
through 100 years, a lot of verification tests have been done and 
it passed all of them starting from weak field solar system 
tests until the recent great discovery of 
gravitational waves from a binary black hole system\cite{Abbott:2016blz}. 
Then, we have entered into the next 100 years for the challenge 
to the discovery of an edge of general relativity. 

Binary pulsar systems, such as the Hulse-Taylor binary\cite{Hulse:1974eb}, 
have been used to test the  gravitational theories 
in the region with a strong gravitational field compared 
to the solar system observations. 
Black hole (BH)-pulsar systems can be also powerful tools for testing the theories, 
and the pulsar can be used as a unique probe for the environment around a BH. 
Future prospects for gravity tests with pulsar-black hole binaries are 
discussed in detail in Ref.~\cite{Liu:2014uka}. 
While they have not been found yet, 
there are indirect evidences of the existence of 
BH-pulsar systems in our galaxy\cite{Eatough:2015jka}. 
One promising system is a pulsar orbiting around ${\rm Sgr~A^*}$ BH. 
Recent simulations indicate that 
200 pulsars exist within 4000~au from ${\rm Sgr~A^*}$
and 
the closest one probably has semi-major axis 120 au\cite{Zhang:2014kva}.
In this paper, taking a pulsar orbiting around ${\rm Sgr~A^*}$ BH into consideration,
we propose an approach to test gravity around a BH with two phenomenological parameters
which characterize deviation from the vacuum Einstein gravity. 
Our phenomenological parameters are different from conventional ones, 
which are used, e.g., in Ref.~\cite{Liu:2014uka}. 

Since the post-Newtonian(PN) approximation is still valid for 
pulsar motion with semi-major axis 120 au, 
we rely on the PN formalism throughout this paper. 
The parametrized post-Newtonian(PPN) formalism is an extension of the PN formalism,
and the most popular phenomenological approach for testing gravitational theories. 
This formalism contains ten free parameters, PPN parameters, 
which appear as coefficients of potentials in the metric
and represent deviation from general relativity(GR). 
Observational constraints on these parameters are summarized in Ref.~\cite{Will:2014kxa}.
If deviation from GR is found in the PPN framework, 
it may support modified gravitational theories. 
However, in the PPN framework, since matter effects are not usually taken into account, 
one may suspect a possible matter effect. 
Then, the story of the Vulcan might be repeated not for 
the solar system but for our galactic center. 
Possibility of unknown matter effects might not be excluded using the PPN framework alone.
Therefore, as a complement to the PPN framework, we
introduce phenomenological parameters 
to see how it deviates from the vacuum case.

In a phenomenological point of view, 
we do not necessarily persist in ordinary healthy matters. 
As in the case of the dark energy problem in cosmology, 
an energy component with an exotic equation of state 
could play a crucial role to explain an actual phenomenon even if 
its origin is not revealed. 
In the case of the dark energy problem, 
there are a lot of attempts to explain the accelerated expansion of our universe 
by using modified gravity theories instead of introducing 
an extra-ordinary matter component. 
In other words, the evidence of the exotic equation of state might imply 
modification of the gravitational theory rather than 
the existence of the exotic matter field. 
Therefore, it is interesting to consider a relativistic exotic matter 
component around a BH with GR to be valid. 
For this purpose, differently from 
the conventional PN formalism, we keep our formalism general enough so that 
relativistic matter components can be treated. 
In analogy with the dark energy problem, 
we introduce an extra-ordinary matter field with the equation of state $p=w\varepsilon$, 
where $p$ and $\varepsilon$ are the pressure 
and energy density, respectively
and $w$ is a constant. 
It should be noted that, although we introduce the extra-ordinary matter component 
in analogy with the dark energy, it is not necessarily identical to 
the dark energy, which causes the accelerated expansion of our universe. 
$\varepsilon$ and $w$ should be regarded as purely phenomenological 
parameters which characterize deviation of the geometry 
from the Kerr BH. 

Pulsars orbiting around ${\rm Sgr~A^*}$ BH can be treated as test particles. 
Therefore we focus on geodesic motion in the spacetime 
described by the PN approximation. 
In the conventional PN scheme, 
the mass density of a fluid contributes to the Newtonian order while
the pressure does only to PN orders. 
However, as is mentioned above, 
we also consider relativistic fluid, in which 
the mass density and the pressure may contribute to the geometry 
at the same order. 
We introduce the fluid component as a small perturbation from 
the vacuum Einstein theory. 
Thus, the mass density and the pressure of the fluid are 
assumed to equally make only post-Newtonian contributions. 
This prescription enable us to describe the geodesic motion around a BH 
with a surrounding relativistic fluid.

This paper is organized as follows. 
In Sec.~\ref{sec:metric} we 
derive the PN metric in the situation of our interest. 
In Sec.~\ref{sec:geodesic}, we focus on the geodesic equation 
and show the difference from it without a fluid component. 
We give an estimate of the effect of the surrounding fluid to 
the pericenter shift in Sec.~\ref{sec:pericenter}. 
Sec.~\ref{sec:SandD} is devoted to a summary and discussion.

In this paper, 
the speed of light and the Newton's gravitational constant are 
denoted by $c$ and $G$, respectively.

\section{Metric}
\label{sec:metric}

In this section, 
we derive the metric of spherically symmetric system 
composed of a BH and a surrounding matter component.
The Einstein equations are given by 
\begin{align}
R^{\mu\nu}-\frac{1}{2}g^{\mu\nu}R=\frac{8\pi G}{c^4}T^{\mu\nu},
\end{align}
where $R^{\mu\nu}$ and $R$ are the Ricci tensor and scalar, respectively, 
$g^{\mu\nu}$ is the spacetime metric 
and $T^{\mu\nu}$ is the energy-momentum tensor.
We consider a central BH and the static spherically symmetric perfect fluid distribution 
having the energy-momentum tensor
\begin{align}
T^{\mu\nu}&=(\rho+\frac{p}{c^2})u^{\mu}u^{\nu}+pg^{\mu\nu},
\label{eq:EMT}
\end{align}
where 
$\rho$ and $p$ are the mass density and the pressure, respectively, 
and $u^{\mu}$ is the four velocity of the fluid element. 
Hereafter, we use a Cartesian coordinate system given by $(t, \bm x)$ 
or $(t, x^j)$, and $r:=|\bm x|$. 
In this notation, the four velocity can be described by 
$u^{\mu}=\gamma(c,\bm{0})$, where 
$\gamma$ is determined by the normalization condition: $g_{\mu\nu}u^{\mu}u^{\nu}=-c^2$.

Since we perform PN expansion, in our approximation, 
the matter density can be described as follows:
\begin{align}
\rho=\frac{\rho^*_{\bullet}(r)}{\gamma \sqrt{-g}}+\frac{\varepsilon}{c^2}
\end{align}
with $\rho^*_{\bullet}(r):=M_\bullet \delta(r)$, 
where 
$M_{\bullet}$ is the mass of BH and $g$ is the determinant of the metric $g_{\mu\nu}$, 
and $\varepsilon$ is the surrounding fluid energy density.
The equation of state of the surrounding fluid is 
assumed to be $p=w\varepsilon$, 
where $w$ is a constant. 
It should be noted that, in a phenomenological point of view, 
global distribution of the fluid is not necessarily needed 
but local distribution near the pulsar trajectory may be enough. 
Therefore, in this paper, we regard the equation of state as an approximate one  
valid only near the pulsar trajectory. 
Then, we do not take the global distribution into account.

Let us introduce expansion parameters.
In the situation of our interest,
a PN expansion parameter $\epsilon$ can be defined by
\begin{align}
\epsilon :=\frac{GM_{\bullet}}{c^2R},
\end{align}
where
$R$ is the reference radius 
given by a characteristic distance scale of the test particle orbit. 
We introduce another expansion parameter defined by 
\begin{align}
\alpha:=\frac{M_R}{M_{\bullet}}:=\frac{4\pi}{3} \frac{({\varepsilon}_R/c^2)R^3}{M_{\bullet}}, 
\end{align}
where 
$\varepsilon_R$ is the energy density of the matter at the radius $R$ from the BH.
We consider that the surrounding fluid is sparse enough 
for $\alpha$ to be regarded as a small quantity. 
Since we have two expansion parameters, for convenience, 
we introduce the following notation $\mathcal{O}(\epsilon^n,\alpha^m)$ 
which denotes higher order terms of 
$\mathcal{O}(\epsilon^n)$ or $\mathcal{O}(\alpha^m)$. 
We consider the geodesic equation up to the order of $\alpha\epsilon$ compared to the 
Newtonian order. 
There is no order $\alpha^2$ term in the geodesic equation. 
Although the order $\epsilon^2$ terms give usual 2PN contributions, 
for simplicity, we neglect them by assuming $\epsilon<\alpha$ in this paper. 
The order $\epsilon^2$ terms can be trivially introduced, and then 
our approach can be extended to the case $\epsilon\sim\alpha$. 
In summary, we neglect $\mathcal{O}(\epsilon^2,\alpha^2)$ terms in the geodesic equation.

Following the method given in Ref.~\cite{Pati:2000I}, 
neglecting the terms proportional to 
$G^3$ in $g_{00}$ and $G^2$ in $g_{jk}$, 
we obtain the following expressions for a spherically symmetric static near-zone metric 
in the standard harmonic gauge:
\begin{align}
g_{00}&=
-1+\frac{2}{c^2}U+\frac{2}{c^4}\big\{3U_p+2P(\rho U)-U^2\big\}
+\frac{4}{c^6}\big\{3P(pU)-3UU_p-P(\rho U_p)\big\}+\mathcal{O}(\epsilon^{3}),
\nonumber \\
g_{jk}&=
\delta_{jk}\Big[1+\frac{2}{c^2}U
+\frac{2}{c^4}\big\{-U_p+2P(\rho U)\big\}\Big]+\mathcal{O}(\epsilon^{2}),\\
g_{0j}&=0,\nonumber
\end{align}
where
\begin{align}
U&:=
G\int\frac{\rho(r')}{|\mbox{\boldmath ${x}$}-\mbox{\boldmath ${x}$}'|}\dd^3x',
\\
U_p&:=
G\int\frac{p(r')}{|\mbox{\boldmath ${x}$}-\mbox{\boldmath ${x}$}'|}\dd^3x',
\\
P(f)&:=
G\int\frac{f(r')}{|\mbox{\boldmath ${x}$}-\mbox{\boldmath ${x}$}'|}\dd^3x'.
\end{align}
Here, the region of integral is the spatial region occupied by the fluid
and $f$ is an arbitrary function.

Decomposing the potential into the BH and fluid parts
and 
using the equation of state for the fluid, 
we obtain 
\begin{align}
g_{00}&=
-1+\frac{2}{c^2}U_{\bullet}+\frac{2}{c^4}\big\{(1+3w)\tilde{U}-U_{\bullet}^2\big\}
\nonumber \\
&\hspace{5mm}
+\frac{2}{c^6}\big\{-2(1+3w)U_{\bullet}\tilde{U}+2(1+3w)P(\varepsilon U_{\bullet})-(1-w)P(\rho_{\bullet}^* \tilde{U})\big\}+\mathcal{O}(\epsilon^{3},\alpha^2), 
\nonumber \\
g_{jk}&=
\delta_{jk}\Big[1+\frac{2}{c^2}U_{\bullet}
+\frac{2}{c^4}(1-w)\tilde{U}\Big]+\mathcal{O}(\epsilon^{2},\alpha^2), 
\end{align}
where the Newtonian potentials of the BH and the fluid are given by
\begin{align}
U_{\bullet}(r)&:=
G\int 
\frac{\rho^*_{\bullet}(r')}
{|\mbox{\boldmath ${x}$}-\mbox{\boldmath ${x}$}'|}\dd^3x'
=\frac{GM_{\bullet}}{r} ,
\\
\tilde{U}(r)&:=
G\int\frac{\varepsilon(r')}{|\mbox{\boldmath ${x}$}-\mbox{\boldmath ${x}$}'|}\dd^3x' .
\end{align}
We set $P(\rho^*_{\bullet}U_{\bullet})=0$ using the regularization prescription: $\delta(r)/r=0$\cite{Gravity2014}, 
which is a special case of the Hadamard regularization.
This prescription yields the same outer metric as that with 
treating the BH as a finite size object.

For later convenience, we consider the following coordinate transformation:
\begin{align}
\bar{x}^j=x^j\big(1-\frac{1}{c^4}A\big)+\mathcal{O}(\epsilon^{3}) ,
\end{align}
where 
$A$ is a constant, which will be determined later. 
The metric after the transformation is 
\begin{align}
g_{00}&=
-1+\frac{2}{c^2}U_{\bullet}+\frac{2}{c^4}\big\{(1+3w)\tilde{U}-U_{\bullet}^2\big\}
\nonumber \\
&\hspace{5mm}
+\frac{2}{c^6}\big\{-2(1+3w)U_{\bullet}\tilde{U}+2(1+3w)P(\varepsilon U_{\bullet})-(1-w)P(\rho_{\bullet}^* \tilde{U})-U_{\bullet}A\big\}+\mathcal{O}(\epsilon^{3},\alpha^2),
\nonumber \\
g_{jk}&=
\delta_{jk}\Big[1+\frac{2}{c^2}U_{\bullet}
+\frac{2}{c^4}\big\{(1-w)\tilde{U}+A\big\}\Big]+\mathcal{O}(\epsilon^{2},\alpha^2). 
\end{align}

Finally we determine the distribution of the perfect fluid 
by solving the Euler equation. 
In our case, it leads to the hydrostatic equilibrium equation:
\begin{align}
&\frac{\dd p(r)}{\dd r}=-\left(\varepsilon(r)+p(r)\right)\frac{GM_{\bullet}}{c^2r^2}
\Leftrightarrow 
\frac{\dd \varepsilon(r)}{\dd r}=-\frac{1+w}{w}\frac{GM_{\bullet}}{c^2r^2}\varepsilon(r) .
\end{align}
Solving the equation with the boundary condition $\varepsilon(R)=\varepsilon_R$, 
we obtain 
\begin{align}
\varepsilon(r)
&=\varepsilon_R\left\{1 +\frac{1+w}{w}\Big(\frac{R}{r}-1\Big)\epsilon
+\mathcal{O}(\epsilon^{2})\right\}.
\label{eq:energy density}
\end{align}
We find that this solution is nonzero at spatial infinity 
unless $\varepsilon_R=0$. 
The near zone metric cannot be defined for the fluid not having a compact support. 
However, as is mentioned before, 
since we consider the equation of state $p=w\varepsilon$ 
is approximately valid only in the vicinity of the test particle orbit, 
we do not care about the distribution beyond the region of our interest. 
Furthermore, 
due to the spherical symmetry, 
the motion of a test particle is independent of the distribution outside the orbit. 
We see this fact in the next section. 
For the same reason, we do not care about the singular behaviour of $\varepsilon$ for 
$r\rightarrow 0$ in this paper. 
In this viewpoint, the parameters $\varepsilon_R$ and $w$ should be regarded as 
just phenomenological parameters which characterize the deviation of 
the local geometry from the vacuum GR case. 
It is worthy to note that, even if we consider these parameters are 
not really matter effects but effective description of some modified gravity theory, 
since the effective energy momentum tensor must be equal to the Einstein tensor, 
it must be compatible with the Bianchi identity. 
Therefore, the hydrostatic equilibrium condition must be, 
at least locally, imposed in any case. 

\section{Geodesic equation}
\label{sec:geodesic}

Let us start with the Lagrangian 
\begin{align}
L
&=-mc\sqrt{-g_{\mu\nu}\frac{dx^{\mu}}{dt}\frac{dx^{\nu}}{dt}} ,
\end{align}
where $m$ is the mass of the test particle.
Expanding this Lagrangian through desired order, 
we obtain 
\begin{align}
L&=
-mc^2\Big[
1
-\frac{1}{c^2}\Big(U_{\bullet}+\frac{1}{2}v^2\Big)
-\frac{1}{c^4}
\Big\{
-\frac{1}{2}U_{\bullet}^2+\frac{3}{2}v^2U_{\bullet}+\frac{1}{8}v^4+(1+3w)\tilde{U}
\Big\}
\nonumber \\
&\hspace{2cm}
-\frac{1}{c^6}
\Big\{
-(1+3w)U_{\bullet} \tilde{U}+\frac{3}{2}\Big(1+\frac{w}{3}\Big)\tilde{U}v^2
+2(1+3w)P(\varepsilon U_{\bullet})
\nonumber \\
&\hspace{6cm}
-(1-w)P(\rho^*_{\bullet}\tilde{U})
-U_{\bullet} A+Av^2
\Big\}
+\mathcal{O}(\epsilon^2,\alpha^2) \Big] ,
\end{align}
where
$v^j=\dd x^j/\dd t$ is the velocity of the test particle
and 
$v^2=\delta_{jk}v^jv^k$.

Euler-Lagrange equations lead to the following geodesic equations:
\begin{align}
\frac{\dd v^j}{\dd t}
&=
\partial_j U_{\bullet}
+\frac{1}{c^2}
\big\{
(v^2-4U_{\bullet})\partial_j U_{\bullet}
-4v^k\partial_k U_{\bullet}v^j
+(1+3w)\partial_j \tilde{U}
\big\} 
\nonumber \\
&\hspace{5mm}
+\frac{1}{c^4}
\big\{
-4(1+3w)U_{\bullet} \partial_j \tilde{U} -4(1+w) \tilde{U}  \partial_j U_{\bullet}
+(1-w)v^2 \partial_j \tilde{U} 
\nonumber \\
&\hspace{5mm}
-4(1+w)v^k\partial_k \tilde{U}v^j
+2(1+3w)\partial_j P(\varepsilon U_{\bullet})
-(1-w)\partial_j P(\rho^*_{\bullet}\tilde{U})
-3A\partial_j U_{\bullet}
\big\}
\nonumber \\
&\hspace{5mm}
+\mathcal{O}(\epsilon^2,\alpha^2) .
\label{eq:geoeq1}
\end{align}
Evaluating these potentials in the geodesic equation 
by using the expression for the fluid energy density \eqref{eq:energy density},
we obtain 
\begin{align}
\frac{\dd \bm{v}}{\dd t}
=-\frac{GM_{\bullet}}{r^2}\bm{n}
&-\frac{1}{c^2}
\Big[
\Big(v^2-4\frac{GM_{\bullet}}{r}\Big)\frac{GM_{\bullet}}{r^2}\bm{n}
-4(\bm{v}\cdot\bm{n})\frac{GM_{\bullet}}{r^2}\bm{v}
+(1+3w)  \frac{4\pi G}{3}r \varepsilon_R \bm{n}
\Big]
\nonumber \\
&-\frac{1}{c^4}
\Big[
\Big\{\frac{3}{2}(1+3w)\frac{1+w}{w}+1-w\Big\} GM_{\bullet} \frac{4 \pi G}{3} \varepsilon_R\bm{n}
\nonumber \\
&\hspace{11mm}
-(1+3w)\frac{1+w}{w} \frac{4 \pi G}{3}r \varepsilon_R c^2 \epsilon\bm{n}
\nonumber \\
&\hspace{11mm}
+(1-w)v^2  \frac{4\pi G}{3}r \varepsilon_R \bm{n}
-4(1+w)(\bm{v} \cdot \bm{n}) \frac{4\pi G}{3}r \varepsilon_R \bm{v}
\nonumber \\
&\hspace{11mm}
-2 \pi G (5+3w) \mathcal{R}^2 \varepsilon_R \frac{GM_{\bullet}}{r^2}\bm{n}
-3A \frac{GM_{\bullet}}{r^2}\bm{n}
\Big]
+\mathcal{O}(\epsilon^2,\alpha^2), 
\label{eq:geoeq}
\end{align}
where
$\bm{n}:=\bm{x}/r$ is the unit vector and 
$\mathcal R$ is the radius of the region with non-zero fluid energy density. 
Here we determine the constant $A$ 
so as to eliminate the term  proportion to $c^{-4}r^{-2}$ in Eq. \eqref{eq:geoeq}
and 
we get 
\begin{align}
A=
-\frac{2}{3} \pi G (5+3w) \mathcal{R}^2 \varepsilon_R.
\end{align}
For this choice of $A$, 
the geodesic equation is independent of $\mathcal{R}$,
which means 
the motion of a test particle is not 
affected by the distribution of the fluid outside the orbit as mentioned before.

Introducing the dimensionless quantities, 
$\bar{\bm{v}}:=\bm{v}/\sqrt{GM/R}$, $\bar{t}:=t/\sqrt{R^3/GM}$ and $\bar{r}:=r/R$,
we obtain
\begin{align}
\frac{\dd \bar{\bm{v}}}{\dd \bar{t}}=
-\frac{1}{\bar{r}^2}\bm{n}
&-\epsilon\Big[
(\bar{v}^2-\frac{4}{\bar{r}})\frac{1}{\bar{r}^2}\bm{n}
-\frac{4\bm{\bar{v}}\cdot\bm{n}}{\bar{r}^2}\bar{\bm{v}}
\Big]
-\alpha (1+3w) \bar{r} \bm{n}
\nonumber \\
&-\alpha \epsilon
\Big[
\Big\{\frac{3}{2}(1+3w)\frac{1+w}{w}+1-w\Big\} \bm{n}
-(1+3w)\frac{1+w}{w} \bar{r} \bm{n}
\nonumber \\
&\hspace{35mm}
+(1-w) \bar{v}^2  \bar{r}  \bm{n}
-4(1+w)(\bar{\bm{v}} \cdot \bm{n}) \bar{r} \bar{\bm{v}}
\Big]
+\mathcal{O}(\epsilon^2,\alpha^2). 
\label{eq:dlgeoeq}
\end{align}
The first term of the equation is the Newtonian term, 
the second one is the 1PN term 
and 
others are the terms describing contributions from the fluid.
The values of expansion parameters are estimated as
\begin{align}
\epsilon&=
10^{-4}
\left(\frac{M_\bullet}{10^6 M_{\odot}}\right) 
\left(\frac{R}{100~{\rm au}}\right)^{-1} ,
\\
\alpha&=
2\times10^{-3} 
\left(\frac{\rho_R}{10^{-9}~{\rm g/cm^3}}\right)
\left(\frac{M_\bullet}{10^6 M_{\odot}}\right)^{-1} 
\left(\frac{R}{100~{\rm au}}\right)^3 ,
\end{align}
where $\rho_R=\varepsilon_R/c^2$ is the mass density of the fluid.
$\alpha$ and $w$ are the phenomenological parameters, 
which describe the deviation from the vacuum GR.

\begin{figure}[htbp]
\begin{center}
\includegraphics[scale=0.73]{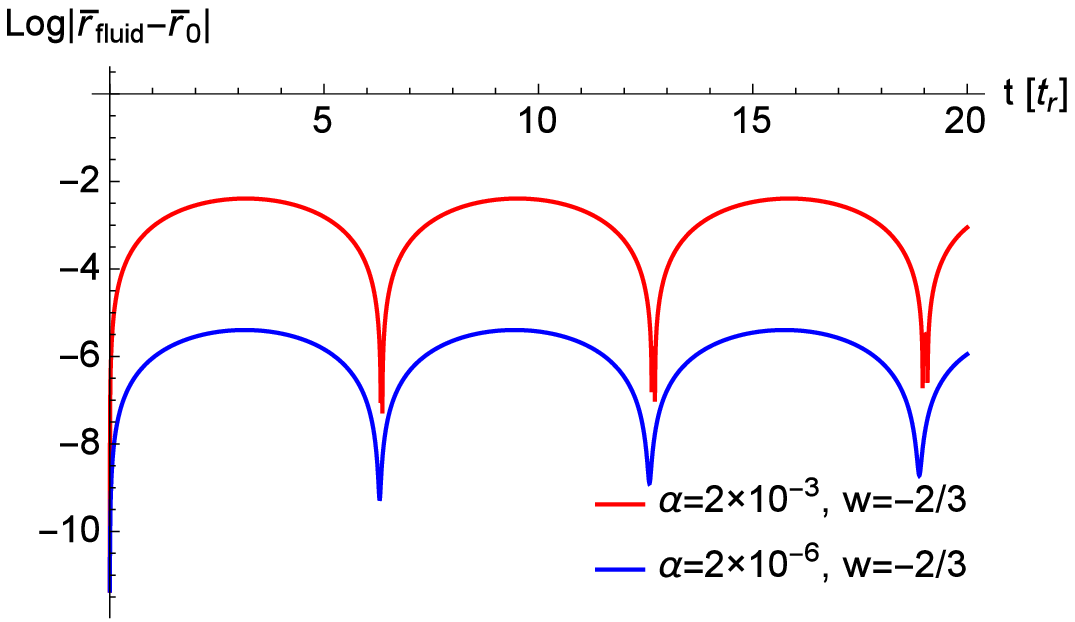}
\includegraphics[scale=0.73]{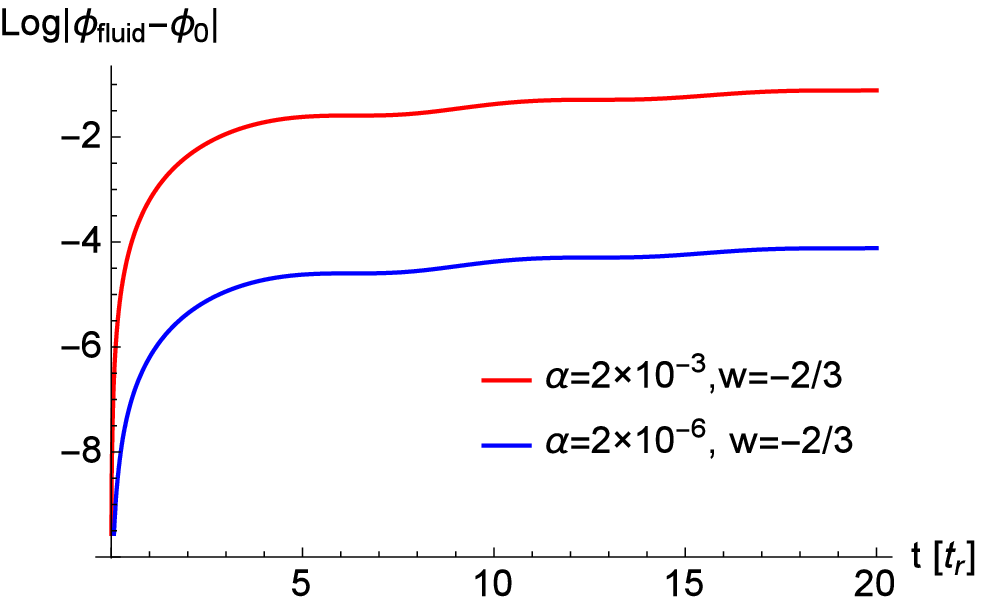}
\caption{
The values of $|\bar{r}_{\rm fluid}-\bar{r}_0|$ and $|\phi_{\rm fluid}-\phi_0|$ 
are plotted as functions of $\bar t$ for the two cases with 
$\rho_R=10^{-9}~{\rm g/cm^3}$ and $\rho_R=10^{-12}~{\rm g/cm^3}$. 
The value of $w$ is set to be $-2/3$. 
}
\label{fig}
\end{center}
\end{figure}

Let us consider how the orbital radius and phase deviate from 
the vacuum GR case. 
In our setting, the orbital motion is restricted within a fixed orbital 
plane as in the Keplerian case. 
The 1PN term and the terms proportional to $\alpha$ 
in the geodesic equation cause deviation from the Keplerian motion. 
The orbital radius of the full system $\bar{r}_{\rm fluid}$ is 
different from that without fluid $\bar{r}_0$. 
The orbital phases $\phi_{\rm fluid}$ and $\phi_0$ 
also differ from each other, 
where the orbital phase is defined by $\phi:=\arctan(y/x)$. 
Eq.~\eqref{eq:dlgeoeq} can be numerically integrated by using 
the {\sf NDSolve} in Mathematica.  
The deviation from the vacuum case is explicitly shown in Fig. \ref{fig} 
for two specific parameter sets. 
%
%
%
%
%
%
%
%
%
The unit time $t_r$ is estimated as 
\begin{align}
t_r=
0.1~{\rm yr}
\left(\frac{M_\bullet}{10^6 M_{\odot}}\right)^{-1/2} 
\left(\frac{R}{100~{\rm au}}\right)^{3/2}. 
\end{align}
We set the initial condition such that 
the orbit is the circle with the radius $R$ in the Newtonian order.
While the difference of the orbital radius is periodic, 
that of the orbital phase monotonically increases. 
We note that the value of $|\bar{r}_{\rm fluid}-\bar{r}_0|$ 
can vanish because the two orbits intersect with each other 
during 1 cycle of the orbit. 
The period of the periodic behaviour is roughly given by 
$2\pi t_r$, which is the period of the Newtonian orbital motion.

\section{pericenter shift}
\label{sec:pericenter}

When we evaluate values of model parameters in a theoretical model 
by using observational signals from a pulsar,  
the parametrized post-Keplerian (PPK) formalism\cite{DD1985,DD1986,Taylor1989,Damour1992} is used. 
In the PPK formalism, deviation from the Kepler motion 
is characterized by the PPK parameters. 
Therefore, we extract the values of 
PPK parameters from observational signals first. 
Then, comparing those observational values with theoretical values, 
we can determine the model parameters. 
In our case, $\rho_R$, $w$ are the model parameters 
characterizing the system.

We roughly estimate the range of $\rho_R$ in which the model
parameters can be determined from the future SKA observation\cite{Liu:2011ae}.
For this purpose, 
we focus on the precession of the pericenter, which is contained in the PPK parameters. 
In our setting, if we take an average for one period, 
change of other Keplerian parameters vanish differently from the effects of 
the spin-orbit coupling(see, e.g. Ref.~\cite{Gravity2014}). 
We use the osculating method(see Appendix \ref{sec:osc}) 
to calculate the pericenter shift due to the surrounding fluid.
Setting $R=a$, 
where $a$ is the semi-major axis of the orbit, 
the pericenter shift due to the existence of the fluid can be 
calculated as 
\begin{align}
\Delta\omega_{\alpha}&=
-3\pi\alpha \sqrt{1-e^2}(1+3w) ,
\label{eq:omegaalpha}
\\
\Delta\omega_{\alpha\epsilon}
&=
5\pi\alpha\epsilon\sqrt{1-e^2}
\Big(1+\frac{11}{5}w\Big) ,
\label{eq:omegaalphaep}
\end{align}
where $e$ is the eccentricity of the orbit, 
and the detailed calculation is given in Appendix \ref{sec:osc}.
Therefore, we obtain 
\begin{align}
\Delta \omega_\alpha &\sim 
-6 \times 10^{-5} \left(1+3w\right) 
\left(\frac{1-e^2}{0.75}\right)^{1/2}
\left(\frac{\rho_R}{10^{-12}~{\rm g/cm^3}}\right)
\left(\frac{M_{\bullet}}{10^6M_\odot}\right)^{-1}
\left(\frac{a}{100~{\rm au}}\right)^{3},
\\
\Delta \omega_{\alpha\epsilon} &\sim 
1 \times 10^{-5}
\Big(1+\frac{11}{5}w\Big)
\left(\frac{1-e^2}{0.75}\right)^{1/2}
\left(\frac{\rho_R}{10^{-9}~{\rm g/cm^3}}\right)
\left(\frac{a}{100~{\rm au}}\right)^{2} .
\end{align}

In order to evaluate detectable values of 
$\Delta \omega_\alpha$ and $\Delta \omega_{\alpha\epsilon}$, 
we compare these values with the effect of BH spin. 
For the BH having a spin vector $\bm{S}$, 
the spin parameter $\chi$ is defined by $\chi:=c|\bm{S}|/GM^2$. 
The spin vector is assumed to be parallel to the angular momentum of the test particle. 
Then, we can calculate the acceleration due to 
the spin effect(see, e.g. Ref.~\cite{Gravity2014}) 
in the same way adopted in Sec. \ref{sec:geodesic}
as follows: 
\begin{align}
\frac{\dd \bar{\bm{v}}}{\dd \bar{t}}=2\epsilon^{3/2}\chi
\frac{(\bar{\bm{v}}\cdot\bm{\lambda})\bm{n}-(\bar{\bm{v}}\cdot\bm{n})\bm{\lambda}}{\bar{r}^3}, 
\end{align}
where $\bm \lambda$ is the unit vector with $\bm{\lambda}\perp \bm{n}, \bm{S}$.

The pericenter shift due to the spin effect $\Delta \omega_{\rm spin}$ is derived 
in the osculating method: 
\begin{align}
\Delta \omega_{\rm spin}&
=-\frac{8\pi}{(1-e^2)^{3/2}}\epsilon^{3/2}\chi 
\nonumber \\
&\sim
-4\times10^{-5} \chi 
\left(\frac{1-e^2}{0.75}\right)^{-3/2}
\left(\frac{M_\bullet}{10^6M_\odot}\right)^{3/2}
\left(\frac{R}{100~{\rm au}}\right)^{-3/2} .
\end{align}
The spin parameter of Sgr $\rm A^*$ could be measured with precision of $\sim0.1\%$ after five years of observations with SKA\cite{Liu:2011ae}. 
Comparing $\Delta \omega_\alpha$ and $\Delta \omega_{\rm spin}$, 
we can conclude that 
if $\rho_R\sim 10^{-12}~{\rm g/cm^3}$, 
the value of $\rho_R(1+3w)$ can  be measured with precision of $\sim 0.1\%$ 
but the value of $\rho_R$ and $w$ cannot be measured independently.
For much denser fluid, 
$\rho_R\sim 10^{-9}~{\rm g/cm^3}$, 
we can expect to measure the value of $\rho_R$ and $w$ with precision of $\sim0.1\%$.
We need to perform detailed simulation as is done in Ref.~\cite{Liu:2011ae} 
to accurately estimate detectability of $\rho_R$ and $w$. 

\section{Summary and Discussions}
\label{sec:SandD}

In this work, 
we have proposed another approach to test the gravity 
around a BH with a surrounding matter component.
We have treated the BH as a point-like mass and considered 
relativistic perfect fluid, 
whose pressure can make the same order contribution to the geometry 
as that from the mass density. 
For simplicity, we have assumed a static spherically symmetric system.
Adopting the PN approximation, we have derived the geodesic equation up to 
the desired order. 
We have estimated the pericenter shift due to the effects of the fluid 
and shown that 
the mass density and the parameter of the equation of state $w$ 
can be determined with precision of $\sim 0.1\%$
if the mass density around the pulsar orbit is $\sim 10^{-9}~{\rm g/cm^3}$. 

Our analysis may be affected by environmental effects around BH\cite{Barausse:2014tra}.  
Significance of the effects due to baryonic gas and stars around BH 
is summarized in Ref.~\cite{Psaltis:2015uza}.
Especially, the perturbation due to stellar distribution 
is discussed in Ref.~\cite{Merritt:2009ex}.
The results in Ref.~\cite{Merritt:2009ex} show that the pericenter 
shift due to the stellar distribution is given by $\Delta \omega_{\alpha}$ with $w=0$. 
Thus, the effect of star distribution can be taken into account within our formulation. 
The dynamical friction from the interstellar gas for pulsar motion is discussed in Ref.~\cite{Psaltis:2011ru}. 
The dynamical friction from the relativistic fluid yields an 
extra advance of pericenter 
and it is order of $(M_{\rm p}/M_{\bullet})\alpha$, 
where $M_{\rm p}$ is the mass of the pulsar and $M_{\rm p}/M_{\bullet}\sim 10^{-6}$.
For an orbit with $\epsilon\sim 10^{-4}$, 
the dynamical friction contribution is two orders of magnitude smaller 
than the contribution of the order of $\alpha\epsilon$. 

If we observe deviation from the vacuum GR, 
there are two possibilities to explain the deviation; 
one is the existence of unknown mater components and 
the other is an alternative theory to GR. 
In our formulation, 
the observed value of $w< 0$ indicates that 
some exotic matter exists around the BH 
with GR to be valid. 
In other words, if $w< 0$, we are confronted with a choice of
accepting the exotic matter or modifying gravitational theory from GR.

It would be interesting to consider 
extension of our formulation to, for example, 
stationary systems, BH-pulsar systems and 
non-spherically symmetric systems. 
They are left as future work.

\section*{Acknowledgements}
We thank K. Takahashi, H. Nakano, K. Yagi and S. Isoyama for helpful comments.

\appendix

\section{Pericenter Shift with the Osculating Method}
\label{sec:osc}

In this Appendix, 
we calculate the pericenter shift based on the usual perturbation scheme 
known as the method of osculating orbital elements\cite{Euler1748, Lagrange1809}.
Here, we consider the terms proportional to $\alpha$ 
in the geodesic equation (\ref{eq:geoeq}) 
as perturbing forces. 
Let us 
start with a general form of the geodesic equation:
\begin{align}
\frac{\dd \bm{v}}{\dd t}&=
-\frac{GM}{r^2}\bm{n}
+\bm{f},
\label{eq:EoM}
\end{align}
where $M$ is the mass of the central object and \bm{f} is a perturbing force 
per unit mass. 
We take the $x$-$y$ plane as orbital plane, 
and the direction of the angular momentum is $z$-direction. 
We introduce base vectors $\bm n:=\bm x/|\bm{x}|$ and $\bm \lambda$ such that 
$\bm{\lambda}\perp \bm{n}, \bm{e}_z$ and $|\bm \lambda|=1$. 
Then, the velocity $\bm v$ and the angular momentum $\bm h$ can be expressed as follows:
\begin{align}
\bm{v}&=(\bm{v}\cdot\bm{n})\bm{n}
+(\bm{v}\cdot\bm{\lambda})\bm{\lambda} ,
\\
\bm{h}:&=\bm{x}\times\bm{v}:=h\bm{e}_z ,
\nonumber 
\end{align}
where 
$h=|\bm{h}|$. 
The perturbing force can be decomposed as 
\begin{align}
\bm{f}=\mathcal{A}\bm{n}+\mathcal{B}\bm{\lambda},
\end{align}
where we have assumed that the force along $\bm{e}_z$ is zero for simplicity.

Through a conventional method(see, e.g. Ref.~\cite{Gravity2014}), 
we can derive the following expression for 
the derivative of the longitude of pericenter $\omega$ with respect to the 
true anomaly $\nu$(the angle between the pericenter and the position vector $\bm x$): 
\begin{align}
\frac{\dd \omega}{\dd \nu}&\simeq
\frac{(1-e^2)^2}{e}\frac{a^2}{GM}
\Big[
-\frac{\cos \nu}{(1+e \cos \nu)^2}\mathcal{A}
+\frac{2+e \cos \nu}{(1+e \cos \nu)^3}\sin \nu~\mathcal{B}
\Big] ,
\end{align}
where $a$ and $e$ are the semi-major axis and the eccentricity, respectively.
We can express this equation in terms of dimensionless quantities:
\begin{align}
\frac{\dd \omega}{\dd \nu}&\simeq
\frac{(1-e^2)^2}{e}
\left(\frac{a}{R}\right)^2
\Big[
-\frac{\cos \nu}{(1+e \cos \nu)^2}\bar{\mathcal{A}}
+\frac{2+e \cos \nu}{(1+e \cos \nu)^3}\sin \nu~\bar{\mathcal{B}}
\Big] ,
\label{eq:domega/dnu}
\end{align}
where 
$\bar{\mathcal{A}}:=\mathcal{A}R^2/(GM)$ and $\bar{\mathcal{B}}:=\mathcal{B}R^2/(GM)$.
The pericenter shift for one period is given by 
\begin{align}
\Delta \omega=\int_0^{2\pi} \frac{\dd \omega}{\dd \nu}\dd \nu.
\label{eq:periadvance}
\end{align}

From Eq. \eqref{eq:dlgeoeq} the components of the perturbing force are expressed as 
\begin{align}
\bar{\mathcal{A}}=&
-\alpha (1+3w) \bar{r}
\nonumber \\
&-\alpha \epsilon
\Big[
\Big\{\frac{3}{2}(1+3w)\frac{1+w}{w}+1-w\Big\} 
-(1+3w)\frac{1+w}{w} \bar{r} 
\nonumber \\
&\hspace{38mm}
+(1-w) \bar{v}^2  \bar{r}  
 -4(1+w)(\bar{\bm{v}} \cdot \bm{n})^2 \bar{r}
\Big] ,
\label{eq:A}
\\
\bar{\mathcal{B}}=&
4\alpha \epsilon (1+w)(\bar{\bm{v}} \cdot \bm{n}) (\bar{\bm{v}} \cdot \bm{\lambda})\bar{r} ,
\label{eq:B}
\end{align}
where the right-hand side is evaluated by using the Kepler relation:
\begin{align}
\bar{r}=\frac{a}{R}\frac{1-e^2}{1+e\cos \nu} ,
\hspace{1cm}
\bar{\bm{v}} \cdot \bm{n}=\sqrt{\frac{R}{a}}\frac{e\sin \nu}{\sqrt{1-e^2}},
\hspace{1cm}
\bar{\bm{v}} \cdot \bm{\lambda}=\sqrt{\frac{R}{a}}\frac{1+e\cos \nu}{\sqrt{1-e^2}} .
\label{eq:3.34'}
\end{align}
Hereafter, for convenience,  we set $a$ as the reference radius $R$. 
From Eq. \eqref{eq:domega/dnu}, \eqref{eq:A} and \eqref{eq:B} we find 
the precession of the pericenter is given by 
\begin{align}
\Big(\frac{\dd \omega}{\dd \nu}\Big)_{\alpha}=&
\alpha(1+3w)
\frac{1}{e}
\frac{(1-e^2)^3 \cos \nu}{(1+e \cos \nu)^3} ,
\\
\Big(\frac{\dd \omega}{\dd \nu}\Big)_{\alpha\epsilon}=&
\alpha\epsilon
\Big[
\Big\{\frac{3}{2}(1+3w)\frac{1+w}{w}+1-w\Big\}
\frac{1}{e}
\frac{(1-e^2)^2 \cos \nu}{(1+e \cos \nu)^2} 
\nonumber \\
&\hspace{8mm}
-(1+3w)\frac{1+w}{w} 
\frac{1}{e}
\frac{(1-e^2)^3 \cos \nu}{(1+e \cos \nu)^3} 
\nonumber \\
&\hspace{8mm}
+(1-w)
\frac{1}{e}
\frac{(1-e^2)^2 \cos \nu}{(1+e \cos \nu)^3} \{2(1+e \cos \nu)-(1-e^2)\}
\nonumber \\
&\hspace{8mm}
+8(1+w) \frac{(1-e^2)^2 \sin^2 \nu}{(1+e \cos \nu)^3} 
\Big].
\end{align}
Substituting these quantities into Eq. \eqref{eq:periadvance}, 
we obtain 
\begin{align}
\Delta\omega_{\alpha}
=&
\alpha(1+3w)I_3(e)
\label{eq:peri-ad-alpha}, 
\\
\Delta\omega_{\alpha\epsilon}
=&
\alpha\epsilon
\Big[
\Big\{\frac{3}{2}(1+3w)\frac{1+w}{w}+3(1-w)-4(1+w)\Big\}I_2(e)
\nonumber \\
&\hspace{3cm}
-\Big\{(1+3w)\frac{1+w}{w} +(1-w)\Big\} I_3(e)
\Big],
\label{eq:peri-ad-alphaep}
\end{align}
where we have defined the following function:
\begin{align}
I_n(e):=\frac{(1-e^2)^n}{e}\int_0^{2\pi} \frac{\cos \nu}{(1+e\cos \nu)^n} \dd \nu.
\end{align}
This function for $n=2$ and $3$ is given by 
\begin{align}
I_2(e)=-2\pi\sqrt{1-e^2},\ \ 
I_3(e)=-3\pi\sqrt{1-e^2}. 
\label{eq:I2I3''}
\end{align}
Therefore, finally we obtain  
\begin{align}
\Delta\omega_{\alpha}&=
-2\pi\alpha 
\sqrt{1-e^2}(1+3w),
\\
\Delta\omega_{\alpha\epsilon}
&=
5\pi\alpha\epsilon\sqrt{1-e^2}
\Big(1+\frac{11}{5}w\Big).
\label{eq:peri-ad}
\end{align}


\end{document}